\documentclass[useAMS,usenatbib]{mn2e}
\usepackage{graphicx}
\usepackage{url}
\usepackage{epstopdf}
\usepackage{color}
\usepackage{bm}
\usepackage{amssymb}
\usepackage{amsmath}
\usepackage{hyperref}
\usepackage{times}
\hypersetup{
    unicode=true,          
    linktocpage=true,
    pdftoolbar=true,        
   pdfmenubar=true,        
    pdffitwindow=true,     
    pdfstartview={FitH},    
    pdfhighlight={/I},
   hyperindex=true,
    pdftitle={My title},    
   pdfauthor={Author},     
   pdfsubject={Subject},   
   pdfcreator={Creator},   
    pdfproducer={Producer}, 
    pdfkeywords={keywords}, 
    pdfnewwindow=true,      
   colorlinks=true,       
    linkcolor=red,          
    citecolor=blue,        
    filecolor=magenta,      
    urlcolor=cyan           
}

\newcommand{\mz}{{\it M-Z}}

\newcommand{\mlimf}{$m_{\mbox{\scriptsize{lim}}}^{\mbox{\scriptsize{f}}}$}
\newcommand{\mlimb}{$m_{\mbox{\scriptsize{lim}}}^{\mbox{\scriptsize{b}}}$}

\newcommand{\mstar}{$m_*$}

\newcommand{\mstarf}{$m_*^{{\mbox{\scriptsize{f}}}}$}
\newcommand{\F}{Figure~\ref}

\newcommand{\tctv}{$T_c$ and $T_v$}
\newcommand{\mlim}{$m_{\mbox{\scriptsize{lim}}}$}
\newcommand{\tc}{$T_c$}
\newcommand{\tv}{$T_v$}
\newcommand{\vmax}{$V/V_{\max}$}
\newcommand{\sn}{{\it s/n}}

\newcommand{\deltazm}{$\delta Z$  and $\delta M$}
\newcommand{\dZ}{$\delta Z$}
\newcommand{\dM}{$\delta M$}

\title[Completeness II. A  signal-to-noise approach for completeness estimators]{Completeness II:  A  signal-to-noise approach for completeness estimators applied to galaxy magnitude-redshift surveys}

\author[Teodoro, Johnston \& Hendry]{ Lu\'{\a i}s Teodoro$^{1,3}\thanks{luis@astro.gla.ac.uk}$, Russell Johnston$^{2,3}\thanks{rjohnston@uwc.ac.za}$ and Martin Hendry$^3\thanks{martin@astro.gla.ac.uk}$
\\
$^1$ELORET Corp., Space Science and Astrobiology Division, MS: 245-3, NASA Ames Research Center, Moffett Field, CA 94035-1000, USA\\
$^2$Department of Physics, University of Western Cape, Belville, Cape Town, South Africa\\
$^3$Department of Physics and Astronomy, Kelvin Building, University of Glasgow, Glasgow, G12 8QQ, Scotland, UK\\
}

\begin{document}

\date{\today}
\pagerange{\pageref{firstpage}--\pageref{lastpage}} \pubyear{2010}

\label{firstpage}

\maketitle

\begin{abstract}
This is the second paper in our completeness series which addresses some of the issues raised in the previous article by \cite{Johnston:2007} in which we developed statistical tests for assessing the completeness in apparent magnitude of magnitude-redshift surveys defined by two flux limits. The
statistics, {\tctv}, associated with these tests are non-parametric and defined in terms of the observed cumulative distribution function of sources; they represent powerful tools for identifying the {\it true} flux limit and/or  characterising systematic errors in magnitude-redshift data.

In this paper we present a new approach to constructing these estimators that resembles an ``adaptive smoothing" procedure -- i.e. by seeking to maintain the same amount the information, as measured by the signal-to-noise ratio, allocated to each galaxy.  For consistency with our previous work, we apply our improved estimators to the Millennium Galaxy Catalogue (MGC) and the Two Degree Field Galaxy Redshift Survey (2dFGRS) data, and demonstrate that one needs to use a {\sn} appropriately tailored for each individual catalogue to optimise the performance of the completeness estimators. Furthermore, unless such an adaptive procedure is employed, the assessment of completeness may result in a spurious outcome if one uses other estimators present in the literature which have not been designed taking into account ``shot noise'' due to sampling.

\end{abstract}

\section{Introduction}
In recent years the statistical analysis of galaxy redshift surveys has played a central role in cosmology, yielding stringent constraints on the parameters of both the underlying cosmological world model and on the clustering properties of galaxies as a function of redshift, environment and morphological type. However, both tasks are hampered by observational selection effects -- due to e.g. detection limits in apparent magnitude, colour, surface brightness or some combination thereof. A wide range of statistical tools has been developed to identify, characterise -- and hopefully to remove -- the impact of observational selection effects from magnitude-redshift surveys.  Presently, we have the initial data release from the  WiggleZ Dark Energy Survey \citep{Drinkwater:2010}, which will attempt to measure the baryon acoustic oscillation (BAO) scale to within 2\% from 240,000 emission line galaxies.   There also has also been the zCOSMOS survey \citep{Lilly:2009,Zucca:2009} that is exploring  galaxy evolution through the role of environment at high redshift in the range $1.5\lesssim z \lesssim3.0$.  To achieve such high precision in these measurements will require accurate understanding of the selection and, particularly with zCOSMOS, luminosity functions.

To fully understand the statistical properties of the aforementioned selection function it is crucial that we understand the role of  completeness in apparent magnitude -- meaning that all galaxies brighter than some specified limiting apparent magnitude (or, as is pertinent to this paper, with apparent magnitudes lying between some specified bright and faint limiting values) have been observed.  A classical test for completeness in apparent magnitude is to analyse the variation in galaxy number counts as a function of the adopted limiting apparent magnitude \citep{Hubble:1926}. This test, which presupposes that the galaxy population does not evolve with time and
is homogeneously distributed in space, is however not very efficient. More specifically, it is difficult to decide in practice whether deviations from the expected galaxy number count are indeed
an effect of incompleteness in apparent magnitude, or are instead due to galaxy clustering and/or evolution of the galaxy luminosity function -- or indeed created by incomplete sampling in apparent magnitude. Of course in designing a completeness test one can also make use of distance information via galaxy redshifts; the still widely used and well-known {\vmax} test of \cite{Schmidt:1968} does this, and considers -- for a specified magnitude limit -- the ratio of two volumes: the volume of a sphere of radius equal to the actual distance of observed galaxy, divided by the volume of a sphere of radius equal to the {\em maximum\/} distance at which the galaxy would be observable -- i.e. at the apparent magnitude limit.  It follows that -- for a non-evolving, homogeneous distribution of galaxies -- the expected value and variance of {\vmax} are equal to $1/2$ and $1/12$ respectively.  The {\vmax} test has been used to assess the completeness of magnitude-redshift samples \citep[see for example][]{Hudson:1991}, but unfortunately it suffers from the same major drawbacks as the Hubble test based on galaxy number counts: it is difficult to interpret whether any significant measured departure from the expected value of {\vmax} is due to incompleteness or to clustering and evolutionary effects.

In a seminal paper, \cite{Efron:1992} (hereafter EP92) introduced a powerful new approach to analysing magnitude-redshift surveys that drew on concepts developed in the so-called $C$-method of \cite{lynden:1971} for constructing galaxy LFs. EP92 proposed a non-parametric permutation test for the independence of the spatial and luminosity distributions of galaxies in a magnitude-limited sample, which required no assumptions concerning the parametric form of both the spatial distribution and the galaxy luminosity function. They applied this test to a quasar sample, with an assumed apparent magnitude limit, in order to robustly estimate the parameters characterising the luminosity distance-redshift relation of the quasars \citep[see also][]{Efron:1999}.

\cite{Rauzy:2001} (hereafter R01) noted that the essential ideas of EP92 could be straightforwardly adapted and extended to turn their non-parametric test of the cosmological model into a non-parametric test of the assumption of a magnitude-limited sample -- thus developing a simple but powerful tool for assessing the magnitude completeness of magnitude-redshift surveys. As was the case with EP -- and unlike the Hubble number counts or {\vmax} tests -- the Rauzy test statistic, {\tc}, requires no assumption about the spatial homogeneity of the galaxy distribution.  Moreover, it also requires no knowledge of the parametric form of the galaxy luminosity function. On the other hand, the Rauzy test was formulated only for the case of a sharp, faint apparent magnitude limit.

\cite{Johnston:2007}  (hereafter JTH07) discussed the advantages of the {\tc}  statistic over
standard completeness tests and extended its use to data that is characterised by both a faint and bright magnitude limit.  Moreover, they introduced a new variant statistic, called {\tv}, constructed using the sampled cumulative distance modulus, $Z$, distribution that retains similar properties to those of {\tc} i.e.  being independent of the spatial distribution of galaxies.  By sampling the data in this way, the {\tv} statistic amounted to a much improved differential version of the the widely used {\vmax} test (which assumes spatial homogeneity).  JTH07 applied their completeness test to three major redshift surveys: the Millennium Galaxy Catalogue (MGC)\citep[e.g.][]{Liske:2003,Cross:2004}, the Two Degree Field Galaxy Redshift Survey (2dFGRS) \citep[e.g.][]{Colless:2001}, and a Sloan Digital Sky Survey - Early Types (SDSS-ET)   \citep[e.g.][]{Bernardi:2003a} sample. They concluded that all three surveys were complete in apparent magnitude up to their respective published magnitude limits. In the case of the 2dFGRS survey data, however, they showed that one is first required to adopt a secondary bright apparent magnitude limit -- i.e. applying the JTH07 generalisation.

Application of the JTH07 generalised completeness test to these three surveys led us to consider two crucial effects that, if not accounted for correctly, could lead to wrong statistical conclusions concerning determination of the {\it true}  completeness limits.   In rough terms, the basic construction of the  {\tctv} statistics  proceeds by identifying  volume-limited subsamples associated with each individual galaxy in the catalogue.  In the design of the orignal Rauzy completeness test, where one is only concerned with the faint apparent magnitude limit, these volume-limited subsamples were uniquely defined and thus could be allowed to grow such that a maximised sampling of the data was achieved.  With the introduction of a secondary bright limit (as shown in \F{Fig:0MZ}) the size of each volume-limited subsample is no longer unique.   This leads to the obvious question: how should one optimally define each subsample?

In studying the distribution of galaxies in the $(M ,Z)$ plane we are seeking to understand the underlying luminosity function of a given population of galaxies, as well as the manner in which that function is sampled. {\it To do so we are, of course, inevitably limited to inferences drawn from a finite number of galaxies.}  This makes the inference process in principle susceptible to shot-noise and thus, if our estimators are constructed from subsamples which are too sparsely populated, might lead to spurious results concerning the global properties of the data-set. In this paper we therefore propose to optimise and extend our current methodology by invoking a well-established and objective criterion: we construct our completeness estimators so as to maximize their local signal-to-noise ratio.

The format of this paper will be as follows.  In \S~\ref{sec:statsframework} we revisit the main points underpinning the construction of the JTH07 {\tctv} statistics.  In \S~\ref{sec:currentissues}  we then explore the adverse consequences that can arise if the JTH07 method is applied without properly accounting for the impact of sparse sampling. For this exploration we use the Millennium Galaxy Catalogue (MGC) and the Two Degree Field Galaxy Redshift Surveys (2dFGRS), as already studied in JTH07, for purely illustrative purposes. This then leads us, in \S~\ref{sec:optimisation}, to propose an optimisation technique that is the first step towards circumventing these issues. In \S~\ref{sec:implementation} we introduce as a sampling threshold a direct measurement of the signal-to-noise ({\sn}) of our sampling technique, and demonstrate how this can be implemented. In \S~\ref{sec:conclusions} we then discuss our conclusions and future work.
\section{The `Separability' Assumption and Statistical Framework}\label{sec:statsframework}
\begin{figure*}\label{Fig:0MZ}
     		\includegraphics[width=0.49\textwidth]{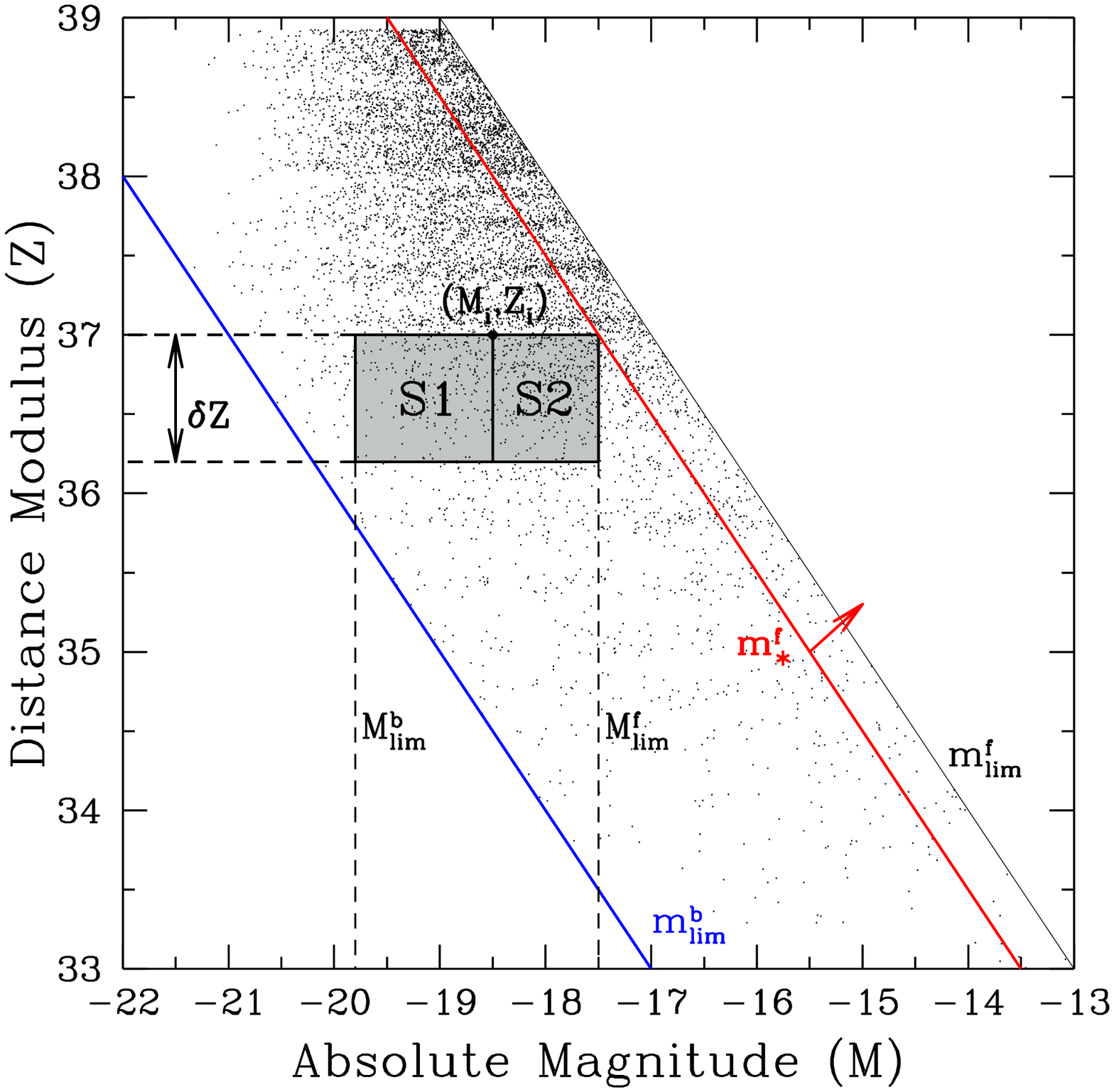}
   		\includegraphics[width=0.49\textwidth]{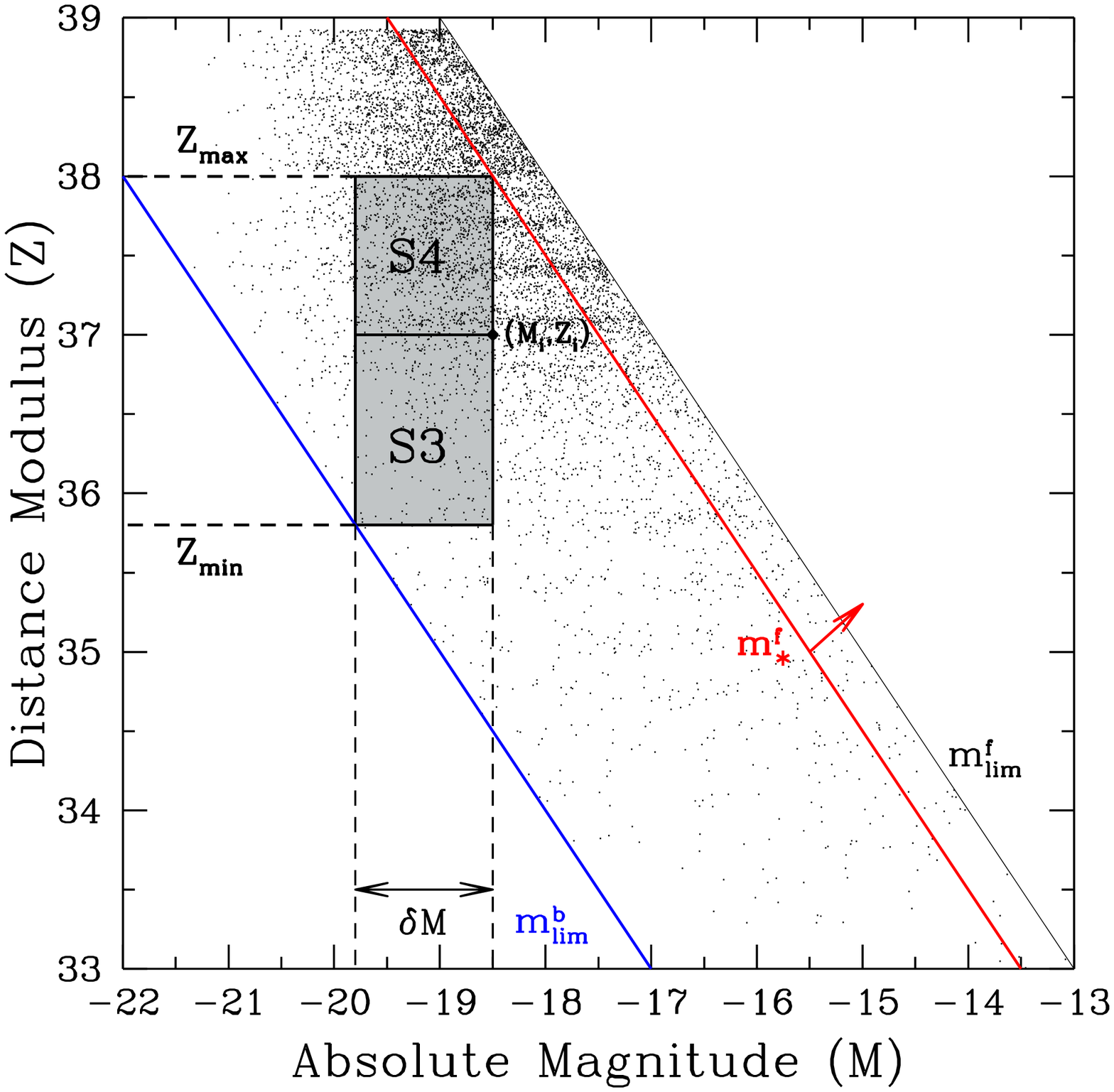}
     \caption{\small Diagram illustrating the construction of the rectangular regions $S_1$, $S_2 $ (left), and $S_3$, $S_4$ (right) which are defined for the random variables, $\zeta_i$ and $\tau_i$ respectively, for a typical galaxy at $(M_i,Z_i)$. The left hand panel shows the construction of the regions $S_1$ and $S_2$  with the inclusion of bright and faint limits {\mlimb}  and {\mlimf}, respectively.   These regions are uniquely defined for a `slice' of specified width, {\dZ}, in distance modulus, and a `trial' faint limit {\mstarf}.  The right hand panel illustrates the construction of the rectangular regions $S_3$ and $S_4$ for $\tau_i$. These regions are uniquely defined for a `slice' of specified width, {\dM}, in absolute magnitude and a `trial' faint limit {\mstarf}.
      }
	\label{Fig:0MZ}
  \end{figure*}
We recall that the fundamental assumption of our method -- also referred to as `separability' -- is that the luminosity function of the galaxy distribution is not dependent on the the three-dimensional redshift space positions {\it{\bf{z}}} $= (z,l,b)$ of the galaxies, where $(l,b)$ are galactic directional coordinates. Although this is a rather restrictive assumption it underlies most of the traditional completeness tests in the literature. The corrected distance modulus $Z$ is defined as,
\begin{equation}
Z \equiv m-M= \mu(z)+k_{\mbox{\scriptsize{corr}}}+e_{\mbox{\scriptsize{corr}}}+A_g(l,b),
\end{equation}
where $k_{\mbox{\scriptsize{corr}}}$ and $e_{\mbox{\scriptsize{corr}}}$ are the $k$-correction and evolutionary correction, respectively, $\mu(z)$ is the distance modulus at redshift $z$ and $A_g(l,b)$ is an extinction correction dependent on galactic coordinates. For simplicity we are marginalising over the galactic directional coordinates.

In assuming separability the joint probability density in absolute magnitude and corrected distance modulus can therefore be written as,
\begin{equation}
\mbox{d}P =  \left[  h(Z)\,\mbox{d}Z \right]\,\left[ f(M)\,dM\right]\,\theta\left({\mbox{\mlimf}} -m\right)\,\theta \left(m-{\mbox{\mlimb}}\right),
\end{equation}
where $f(M)$ and $h(Z)$ are the probability density function of $M$ and $Z$, respectively, and $\theta$ is the Heaviside or `step' function defined as,
\begin{equation}
\theta(x) \equiv  \left \{
\begin{array}{lll}
1 & \mbox{if}  & x\ge 0,\vspace{1.5mm}\\
0 & \mbox{if} & x < 0.
\end{array}
\right .
\end{equation}
Thus for each object $i$ present in a catalogue we define
the random variables $\zeta_i$ and $\tau_i$ for the statistics {\tctv} respectively\footnote{Briefly, {\tctv} are defined as
\begin{displaymath}T_c = \sum_{i=1}^{N_{gal}}\frac{\zeta_i-1/2}{\left[\mbox{Var}(\zeta_i)\right]^{1/2}},\quad \mbox{and}\quad T_v = \sum_{i=1}^{N_{gal}}\frac{\tau_i-1/2}{\left[\mbox{Var}(\tau_i)\right]^{1/2}}, \nonumber \end{displaymath} respectively.}
(for a detailed discussion see JTH07),
\begin{eqnarray}\label{Eq:zetav1}
\zeta_i &=& \frac{F(M_i) - F [ M^{\rm b}_{\rm lim}(Z_i - \delta Z) ]}{F [M^{\rm f}_{\rm lim}(Z_i) ] - F [ M^{\rm b}_{\rm lim}(Z_i - \delta Z) ]} \nonumber \\
&= & \frac{n({S_1})}{n({S_1\cup S_2})}=\frac{r_i}{n_i+1},
\end{eqnarray}
and
\begin{eqnarray}\label{Eq:tauv1}
\tau_i &=&\frac{H(Z_i) - H[Z^{\mbox{\scriptsize{b}}}_{\rm min}(M_i-\delta M)]}{H[Z^{\mbox{\scriptsize{f}}}_{\rm max}(M_i)] - H[Z^{\mbox{\scriptsize{b}}}_{\rm min}(M_i-\delta M)]} \nonumber \\ & =&  \frac{n({S_3})}{n({S_3\cup S_4})}=\frac{q_i}{t_i+1},
\end{eqnarray}
where $r_i$ denotes the number of galaxies belonging to region $S_1$, $n_i$ the number of galaxies belonging to $S_1\,\cup\,S_2$, $q_i$ the number of galaxies belonging to $S_3$, and  $t_i$ the number of galaxies belonging to $S_3\,\cup\,S_4$.  Figure \ref{Fig:0MZ} illustrates the construction of the rectangular regions $S_1$, $S_2$, $S_3$ and $S_4$ as well as the meaning and definition of the slices in magnitude, {\dZ}, and distance modulus, $\delta M$. It should be mentioned that $r_i$ was also the notation used in EP92 to denote the {\it rank} of  the object $i$ when galaxies are sorted by magnitude.

Essentially, the key to the JTH07 extension lay in the introduction of these fixed `slice' widths {\dZ} for $\zeta_i$ and {\dM} for $\tau_i$. Fixing these widths to a predetermined value allows the construction of unique, separable regions in Equations~{\ref{Eq:zetav1} and \ref{Eq:tauv1} within any doubly truncated survey i.e.  for a survey with well defined bright and faint apparent magnitude limits.

However, the choice of the  {\deltazm}  widths is essentially arbitrary, and one might wish to consider  applying different `trial' widths depending on the properties of the data set under study.  JTH07 briefly discussed this point, and noted that by varying the widths in this manner two distinct effects for the determination of the true {\mlimf} were revealed:
\begin{itemize}
\item{For  very  small values of  {\dZ} and {\dM} the respective {\tctv} statistics will be dominated by what we may term `shot-noise' (since the rectangular regions they identify are extremely sparsely sampled); this makes the process of drawing significant conclusions regarding nature of  the true faint apparent magnitude limit impossible.}
\item{Conversely, when the values of  {\deltazm} are taken to be very large, then for data-sets that are not well described by a sharp {\mlim} one appears to observe a range of possible values for the {\em true\/} faint magnitude limit.}
\end{itemize}
We will illustrate in more detail the manifestation of these two effects in the following section.
\section{consequences of sparse sampling}\label{sec:currentissues}
For continuity (and illustrative purposes)  we revisit  the Millennium Galaxy Catalogue (MGC), the Two Degree Field Galaxy Redshift Survey (2dFGRS) and the Sload Digital Sky Survey Early Types (SDSS-ET) samples as used previously  in JTH07.  Please refer to this paper for survey  description and sample selection.
\subsection{`Shot-noise' dominated sampling}
In this section we examine more closely the consequences  of sparse sampling issues in the construction of the random variables, $\zeta_i$ and $\tau_i$, for the statistics {\tctv} respectively.

In  Figure~\ref{Fig:2df001} we have applied the JTH07 {\tctv} estimators  to the SDSS-ET (upper panel),  MGC (middle panel) and 2dFGRS  (lower panel) for selected values of {\deltazm}. (Both {\dZ} and {\dM} are defined in {\F{Fig:0MZ}).  For the SDSS-ET  we observe that the {\tctv} curves corresponding to respective widths of {\deltazm} = 0.001 and 0.01 fluctuate within the $|3\sigma|$ limits for each $m_*$ between the survey  limits of $14.5 <m_{\lim}<17.45$ (as one would expect for a complete sample, following EP92, R01 and JTH07).  However, contrary to the expectations of those earlier papers, as $m_*$ moves beyond the published faint limit of the survey, the {\tctv} curves  drop slightly and then flatten (or `flat-line') inside $-3\sigma<T_c, T_v<3\sigma$ regions, instead of dropping sharply below the $-3\sigma$ level.  Similar results are seen with  MGC at  {\deltazm} = 0.01 and 2dF up to  {\deltazm} $\approx$ 0.02.
\begin{figure*}\label{Fig:2df001}
     \includegraphics[width=0.8\textwidth]{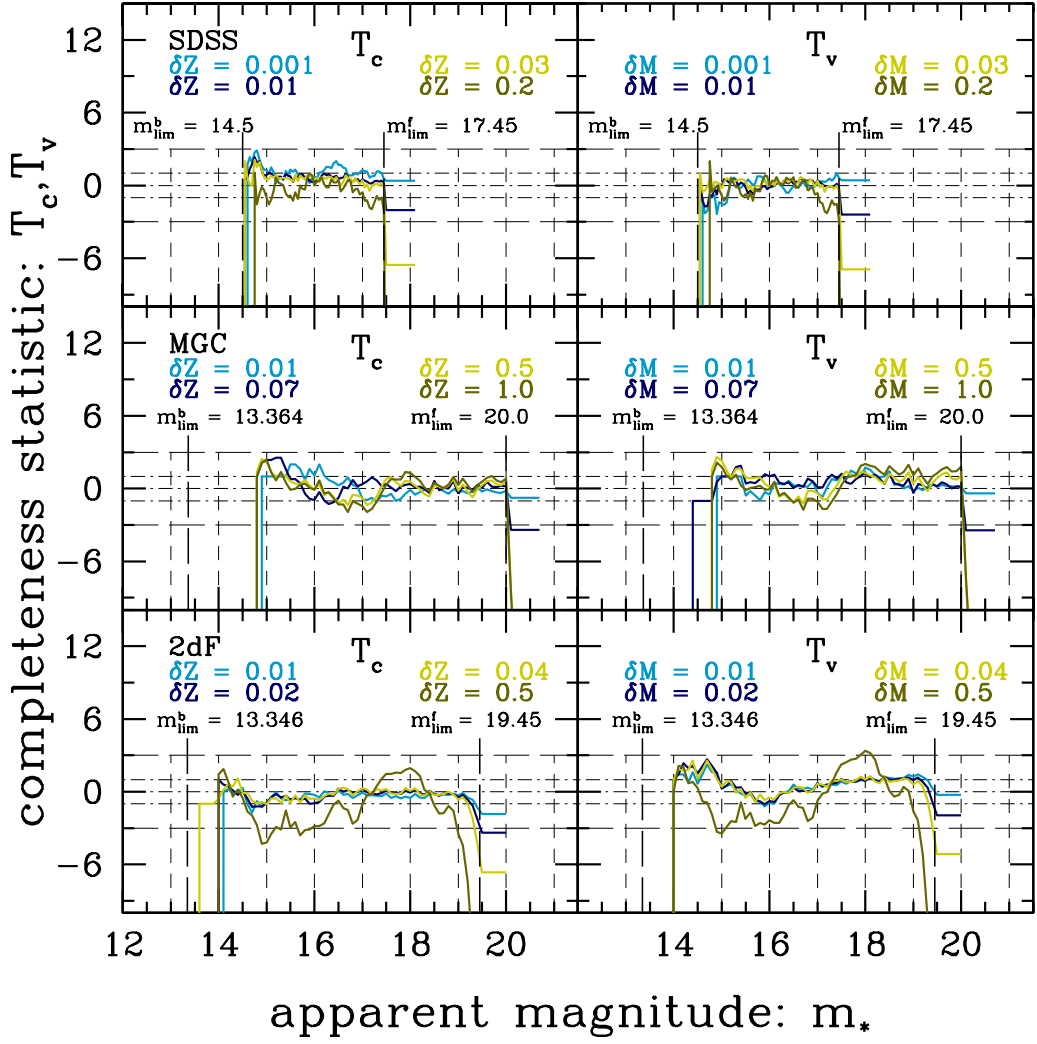}
      \caption{\small {\tctv} Results for  the SDSS-ET  (upper panels), MGC (middle panels) and 2dF (lower panels) applying the JTH07 method  for varying  values of {\deltazm},  where we can observe the transition between `shot-noise' dominated sampling and signal dominated sampling.  We can define this transition to occur at the point where the width of {\dZ} or {\dM}, for {\tctv} respectively, is sufficiently large that the appropriate statistic drops to the $-3\sigma$ level at the faint magnitude limit of the survey.  This occurs at values of  {\deltazm} $\gtrsim$ 0.01 for SDSS, $\gtrsim$ 0.07 for MGC and $\gtrsim$ 0.02 for 2dF. }
	\label{Fig:2df001}
  \end{figure*}
As we move to increasingly larger values of {\deltazm}, as shown in  Figure~\ref{Fig:2df001}, the {\tctv} curves continue to `flat-line' beyond the magnitude limit, but now do so at a value of the statistic which lies increasingly below $-3\sigma$.
\begin{figure*}\label{Fig:2dfdelta}
     \includegraphics[trim=0mm 0mm 0mm 90mm, clip=true,width=0.49\textwidth]{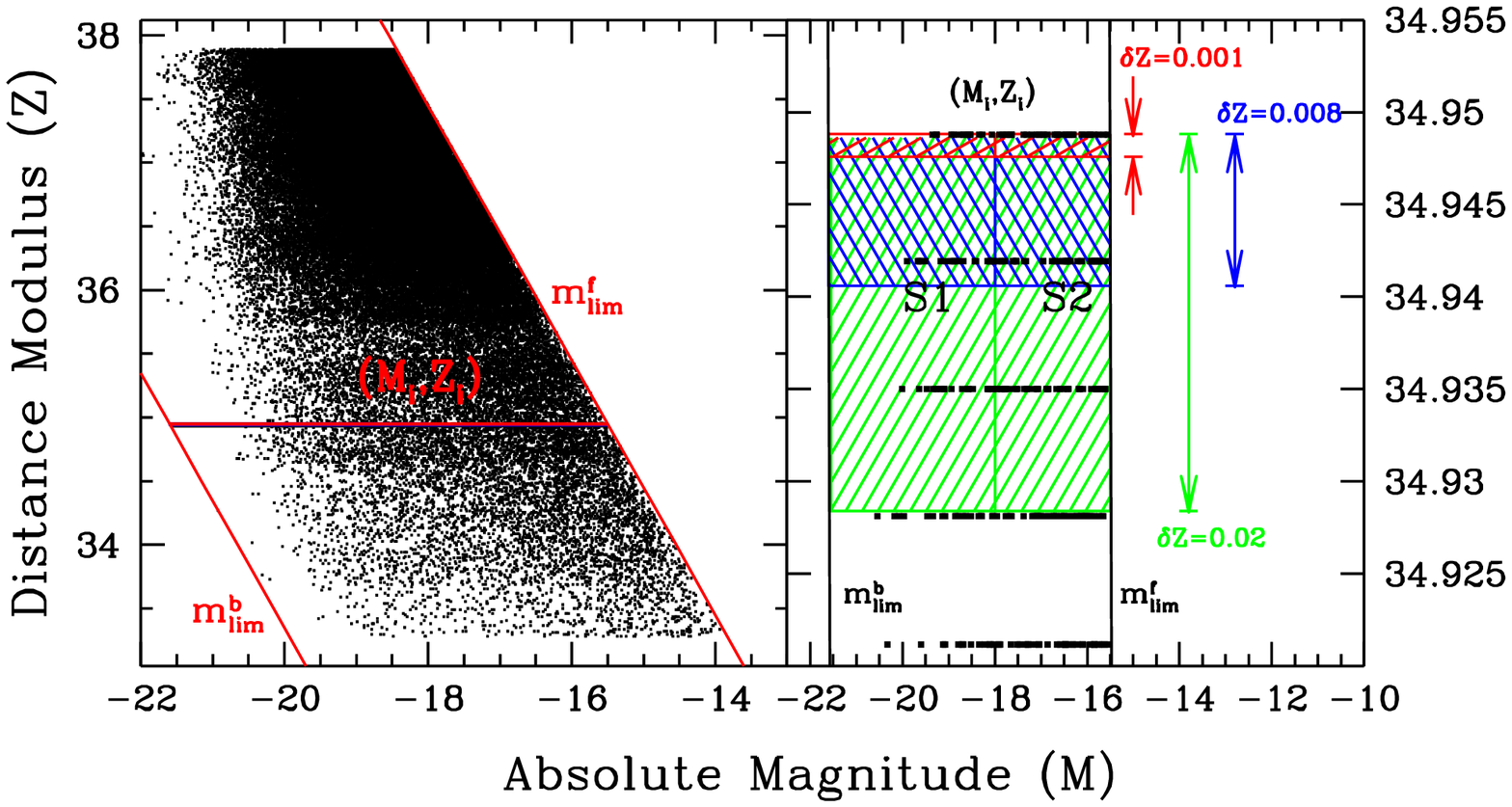} \hfill
     \includegraphics[trim=0mm 0mm 0mm 90mm,clip=true,width=0.49\textwidth]{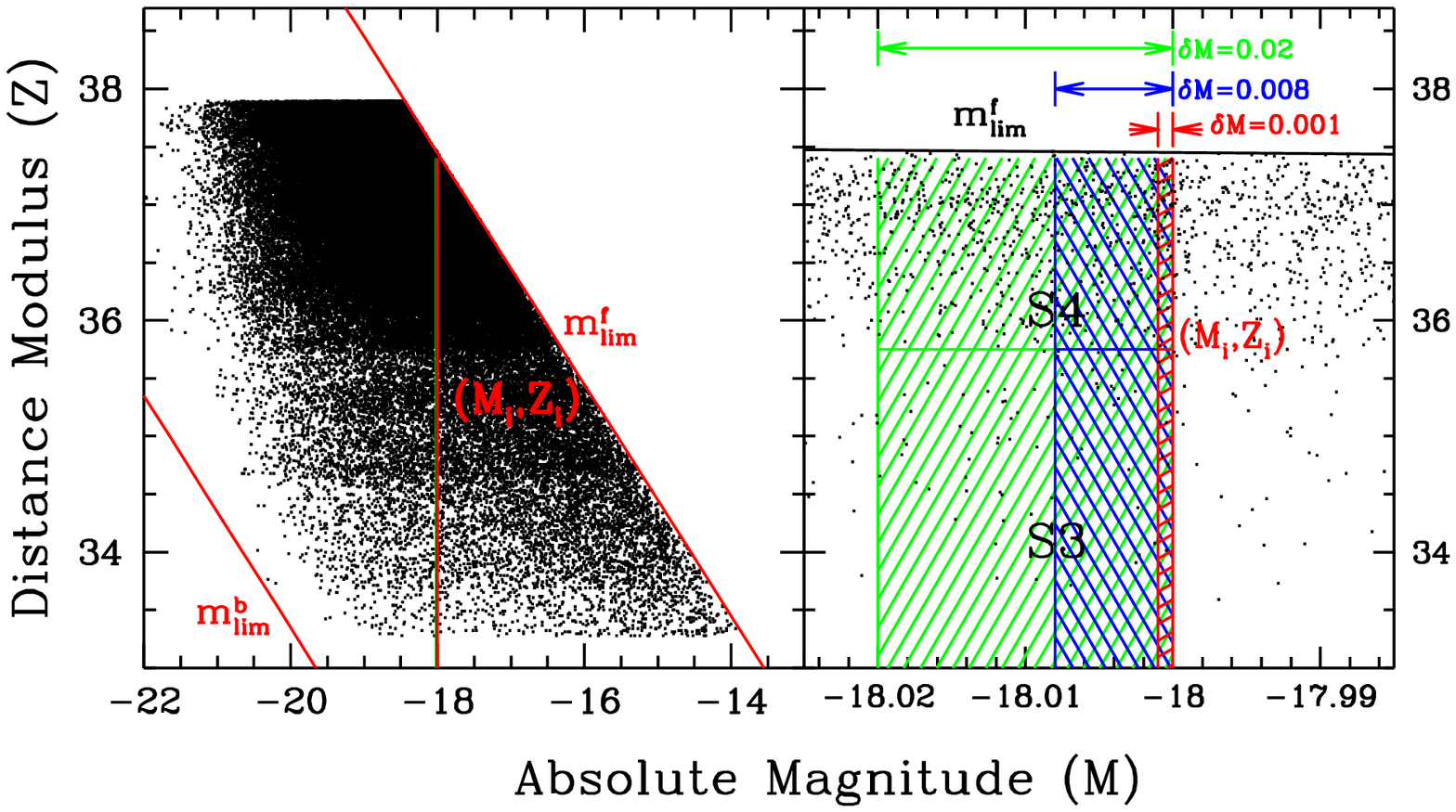}
      \caption{\small Schematic illustrating the cause of the `flat-lining' effect (within the $|3\sigma|$ confidence limits) observed for small values of {\deltazm}.  The left hand panel shows the $(M,Z)$ distribution for the 2dFGRS with the faint apparent magnitude limit {\mlimf} and our adopted bright limit, {\mlimb} indicated as red diagonal lines.  The left hand plot considers the $\zeta_i$ construction for a galaxy at $(M_i, Z_i)$, with {\dZ} = 0.001, 0.008 and 0.02.  The right hand plot zooms in to allow us to see the three distinct regions created as the size of {\dZ}, and the relative number of galaxies contained therein, increases.  Similarly, the right hand panel shows the corresponding $\tau_i$ construction  for a galaxy at $(M_i, Z_i)$, for {\dM} = 0.001, 0.008 and 0.02.}
	\label{Fig:2dfdelta}
  \end{figure*}

This so-called `flat-lining' effect can essentially be used as a means of identifying the `shot-noise' level for a given width of  {\deltazm} -- i.e. the value of {\deltazm} less than which the sampling becomes too sparse to allow the magnitude limit to be reliably estimated. Understanding precisely why this `flat-lining' happens becomes quite straightforward when one considers carefully what are the contributing factors: the number of objects in the catalogue and the range in apparent magnitude of the survey. The effect is illustrated in detail in Figure~\ref{Fig:2dfdelta}.   The left panel shows the now familiar $M$-$Z$ distribution with the red diagonal lines representing the faint apparent magnitude limit {\mlimf}  and our adopted bright limit {\mlimb}.    The main feature of this plot is the narrow red, blue and green rectangles which actually delineate the {\tc} regions $S_1$ and $S_2$ for a galaxy at $(M_i, Z_i)$ with  {\dZ} = 0.001, 0.008 and 0.02 respectively. (Here we are considering a trial $m_*$ equal to the survey limit i.e. {\mlimf} = 19.45~mag.).  Since these rectangular `strips' represent such a tiny fraction of the $M$-$Z$ distribution they can barely be separated in the main diagram. The left panel, therefore, also shows a close-up of this particular region, where the distinctive coloured areas are now clearly defined.   (Note that, because of the very narrow range of distance moduli considered in this close-up, the apparent magnitude limit appears essentially as a vertical line).  The right panel of Figure~\ref{Fig:2dfdelta} represents, for the same galaxy at $(M_i, Z_i)$, the equivalent {\tv} construction with {\dM} = 0.001, 0.008 and 0.02 respectively -- with again the different coloured regions also shown in extreme close-up.

What is immediately apparent for both the {\tctv} statistics is the very small number of galaxies that populate the rectangular regions for these small values of  {\deltazm}. In particular, it is clear that
as $m_*$ is increased beyond the true value {\mlimf}, {\em no\/} further galaxies will be added to the subsets $S_2$ (for $T_C$) and $S_4$ (for $T_V$).  By considering  Equations~\ref{Eq:zetav1} and \ref{Eq:tauv1}, it then follows  that the {\tctv} statistics will remain constant for larger values of $m_*$ -- which explains the `flat-lining' effect seen in Figure~\ref{Fig:2df001}.

The pattern which was apparent in Figure~\ref{Fig:2df001}, whereby the `flat-lining' effect occurred at progressively {\em lower\/} values of {\tctv} as the widths of {\deltazm} were increased, can be extended to the limiting case that corresponds to the original Rauzy (R01) completeness test -- where the absence of a bright apparent magnitude limit means that there is in principle no limit to the height of the constructed regions. However, since we are dealing with a flux-limited catalogue that contains a finite number of galaxies, we can expect that {\em ultimately\/} the `flat-lining' effect will become apparent for the R01 completeness test too, if we consider a sufficiently faint trial value of $m_*$.  This effect is indeed seen in Figure~\ref{Fig:flatline}, albeit for a value of {\tctv} that lies enormously below the characteristic $3-\sigma$ level which one might choose to identify as the value of the statistic indicating the true apparent magnitude limit.

In summary, then, we can understand the `flat-lining' effect as a direct consequence of the very sparse sampling which occurs for small values of {\deltazm}.  A suitable choice for the width of {\deltazm} can then be taken to be the values for which the onset of the `flat-lining' effect {\em only\/} occurs when
the test statistics {\tctv} have already dropped to $3-\sigma$ below their expected value, when the trial apparent magnitude limit is equal to the true value {\mlimf}.

 \begin{figure*}\label{Fig:flatline}
    \begin{center}
     \includegraphics[trim=0mm 60mm 0mm 50mm, clip=true,width=1.\textwidth]{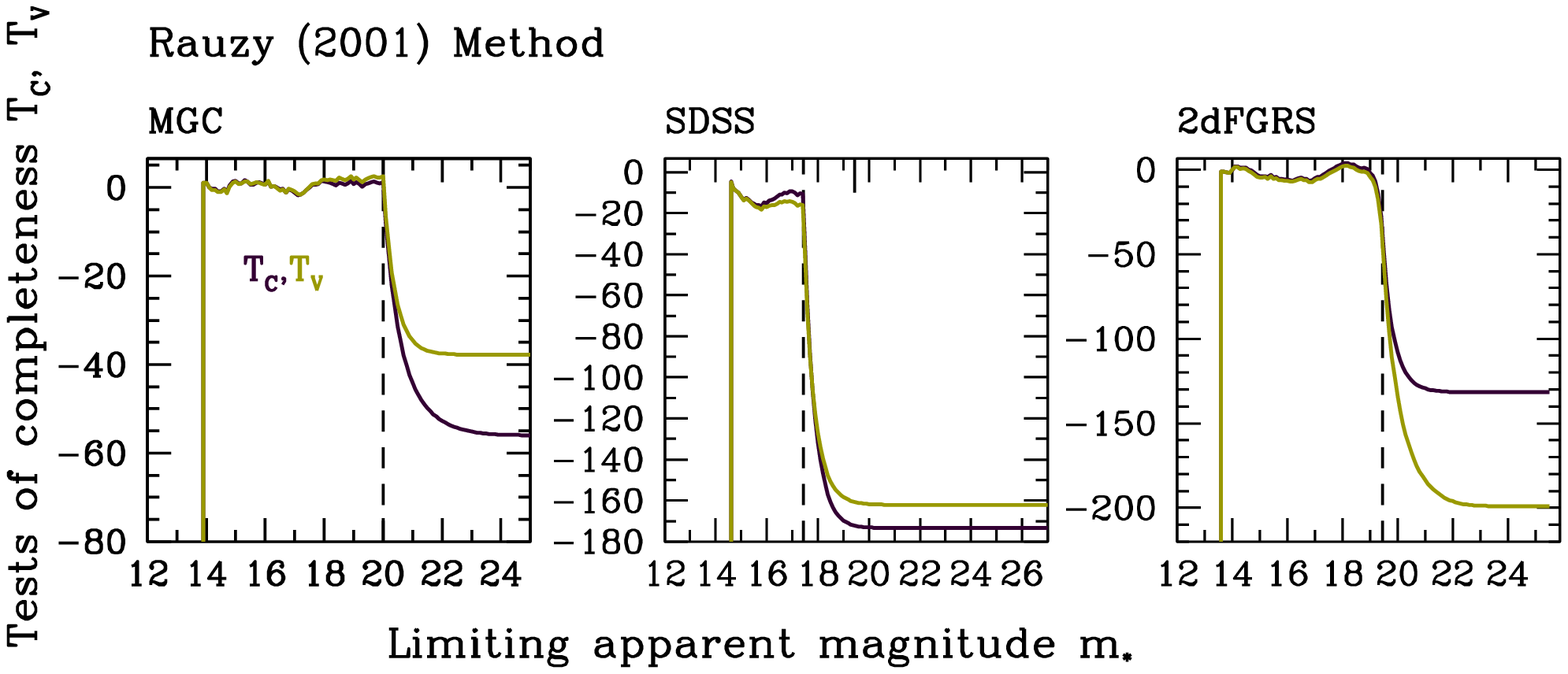}
      \caption{\small {\tctv} plots for 2dFGRS (left) and MGC (right).  Here we apply the R01 method, where the rectangular regions in {\F{Fig:0MZ}} are allowed to grow to their maximum size when accounting for {\mlimf} {\it only\/}.  We can see in both panels that if one allows $m_*$ to pass far enough beyond the magnitude limit of the survey, the `flat-lining' effect will eventually dominate, albeit for extremely negative values of {\tctv}.}
	\label{Fig:flatline}
    \end{center}
  \end{figure*}
\subsection{Variation in $m_{\rm lim}$}
We now briefly consider the apparent variation in the value of $m_{\rm lim}$ determined, resulting from the adoption of larger values of {\deltazm} for a survey that is doubly truncated by a bright and faint magnitude limit.   If we consider once again  {\F{Fig:2df001}} we can see that for both SDSS-ET and MGC as we move to larger values of {\deltazm} our ability to determine correctly the true completeness limit of the survey is unaffected. However, we can quite clearly see in the case of the 2dFGRS on the lower panel that, as  {\deltazm} increases to, and beyond, the point at which the test statistics systematically fall below their $-3\sigma$ level  (which we adopt to indicate the true apparent magnitude limit), the value of  $m_{\lim}^{\rm true}$ also varies with the values of {\deltazm} adopted.     In the range of values we are considering in this example,  $0.002\lesssim\delta Z,\delta M < 0.5$ we actually observe a corresponding range of {\mlim} from $19.0\lesssim m_{\rm lim} \lesssim 19.4$.

This variation in the `true' magnitude limit inferred for a survey is rather unsatisfactory, and would somewhat defeat the purpose of the original Rauzy completeness test: to provide a robust, non-parametric and objective method for independently validating the magnitude completeness of a given survey.  It underlines the importance of optimising the performance of our test statistics -- an issue which we consider in more detail in the following sections.
\section{Expressions for the signal-to-noise of our estimators}\label{sec:optimisation}
In this section we now consider how the estimators $\zeta_i$ and $\tau_i$ are constructed, and in particular how they will be affected by random sampling fluctuations, in order to gain insight on how they might be optimised. This will essentially involve computing a measure of the {\sn} on the sampled $\zeta_i$ and $\tau_i$, and how those variables are affected by fluctuations in the number of galaxies sampled in the regions $S_1$ to $S_4$.
For the moment let us consider $\zeta_i$ only.  If we assume that the survey galaxies are sampled according to a Poisson distribution then we can derive an expression for the Poisson (or shot) noise associated with $\zeta_i$ by applying simple perturbation theory.  In this case  Equation~\ref{Eq:zetav1} then becomes
\begin{equation}\label{Eq:zetanoise1}
\delta \zeta _i = \frac{{\delta r_i (n_i   + 1) - r_i \delta (n_i   + 1)}}
{{(n_i  + 1)^2 }}.
\end{equation}
To take into account the cross-terms we square Equation~\ref{Eq:zetanoise1} to get,
\begin{equation}
(\delta \zeta _i )^2  = \frac{{\delta r_i ^2 }}
{{(n_i   + 1)^2 }} + \frac{{\zeta_i ^2 [\delta (n_i   + 1)]^2 r_i^2 }}
{{(n_i   + 1)^2 }} - \frac{{2\zeta \delta n_i [\delta (n_i   + 1)]}}
{{(n_i   + 1)^2 }},
\end{equation}
and
\begin{equation}\label{Eq:zetasigtonoise}
\frac{{\zeta_i ^2 }}{{(\delta \zeta_i )^2 }}
= \frac{{r_i^2 }}{{\delta r_i ^2 }}
+ \frac{{(n_i   + 1)^2 }}{{[\delta (n_i  + 1)]^2 }}
- \frac{{r_i(n_i   + 1)}}{{2\delta r_i [\delta (n_i   + 1)]}}.
\end{equation}
By applying a similar approach for {\tv} we can obtain a similar expression for the {\sn} associated with estimating $\tau_i$.  Starting from Equation~\ref{Eq:tauv1} we can show that,
\begin{equation}\label{Eq:tausigtonoise}
\frac{{\tau_i}}{{(\delta \tau_i)}}
 = \left [ \frac{{q_i^2 }}{{\delta q_i ^2 }}
 + \frac{{(t_i   + 1)^2 }}{{[\delta (t_i  + 1)]^2 }}
 - \frac{{q_i(t_i   + 1)}}{{2\delta q_i [\delta (t_i   + 1)]}}\right ]^{1/2}.
\end{equation}

%
\begin{figure*}\label{Fig:sn_thresh}
     \includegraphics[width=0.33\textwidth]{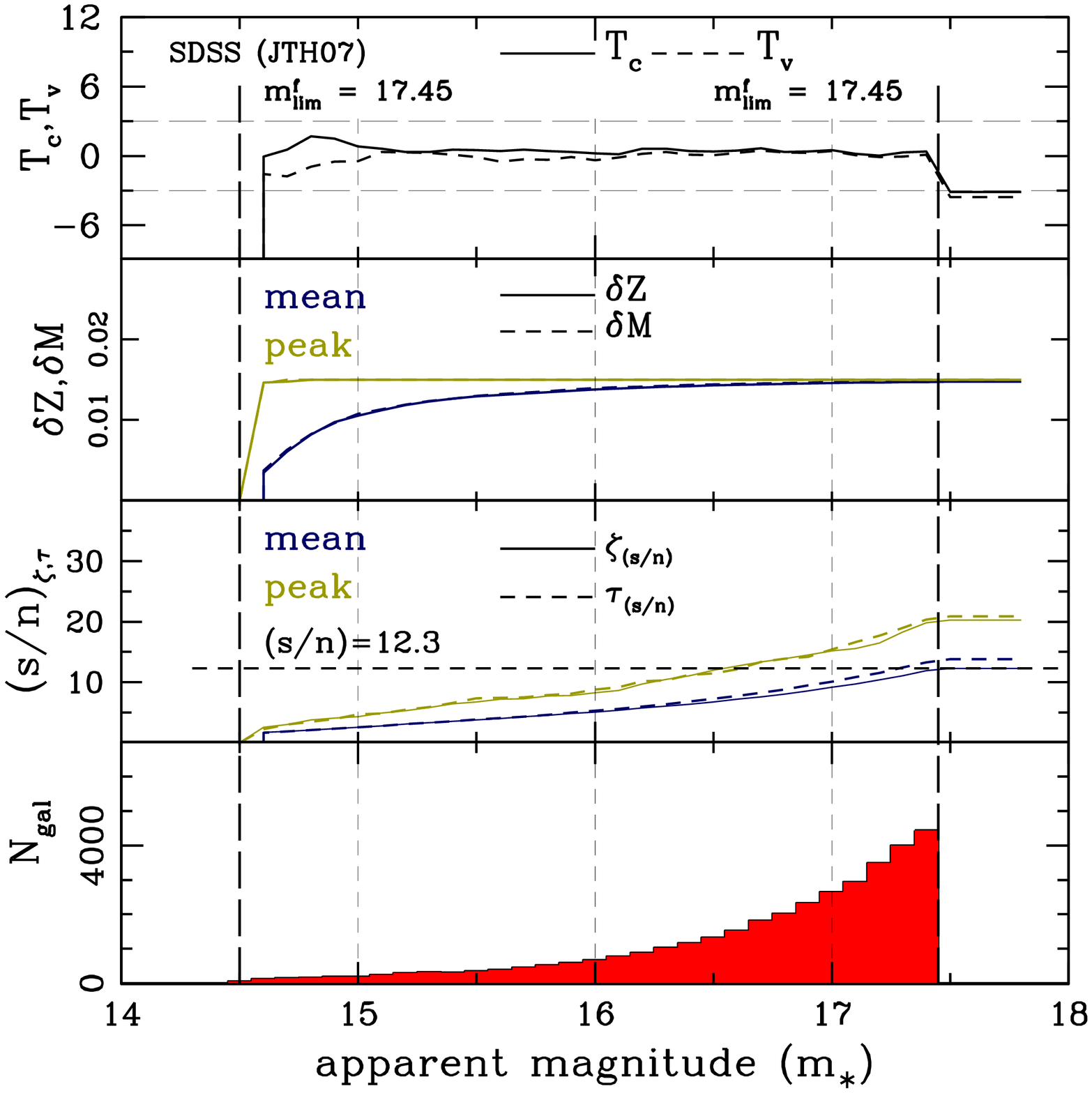}
     \includegraphics[width=0.33\textwidth]{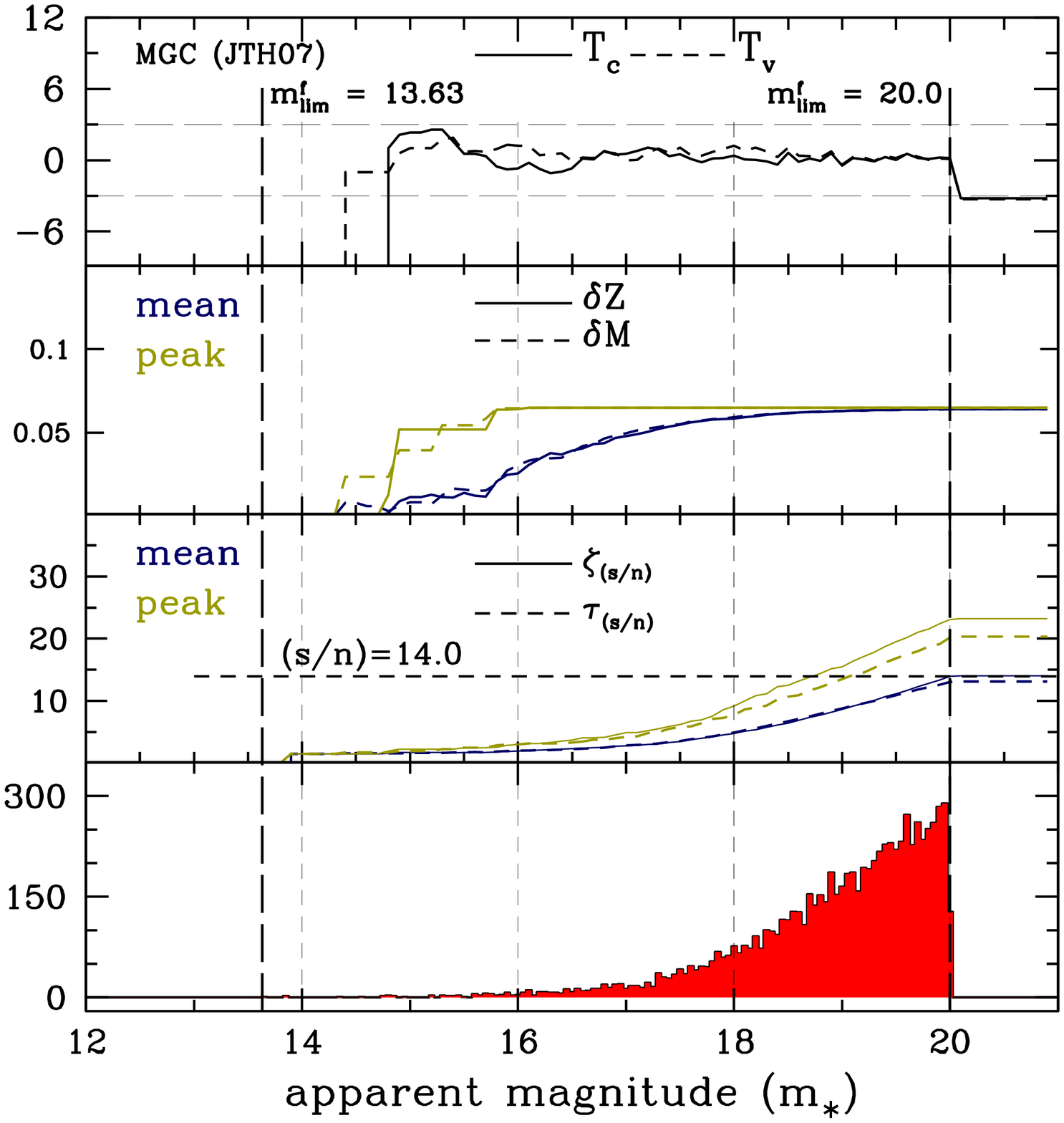}
    \includegraphics[width=0.33\textwidth]{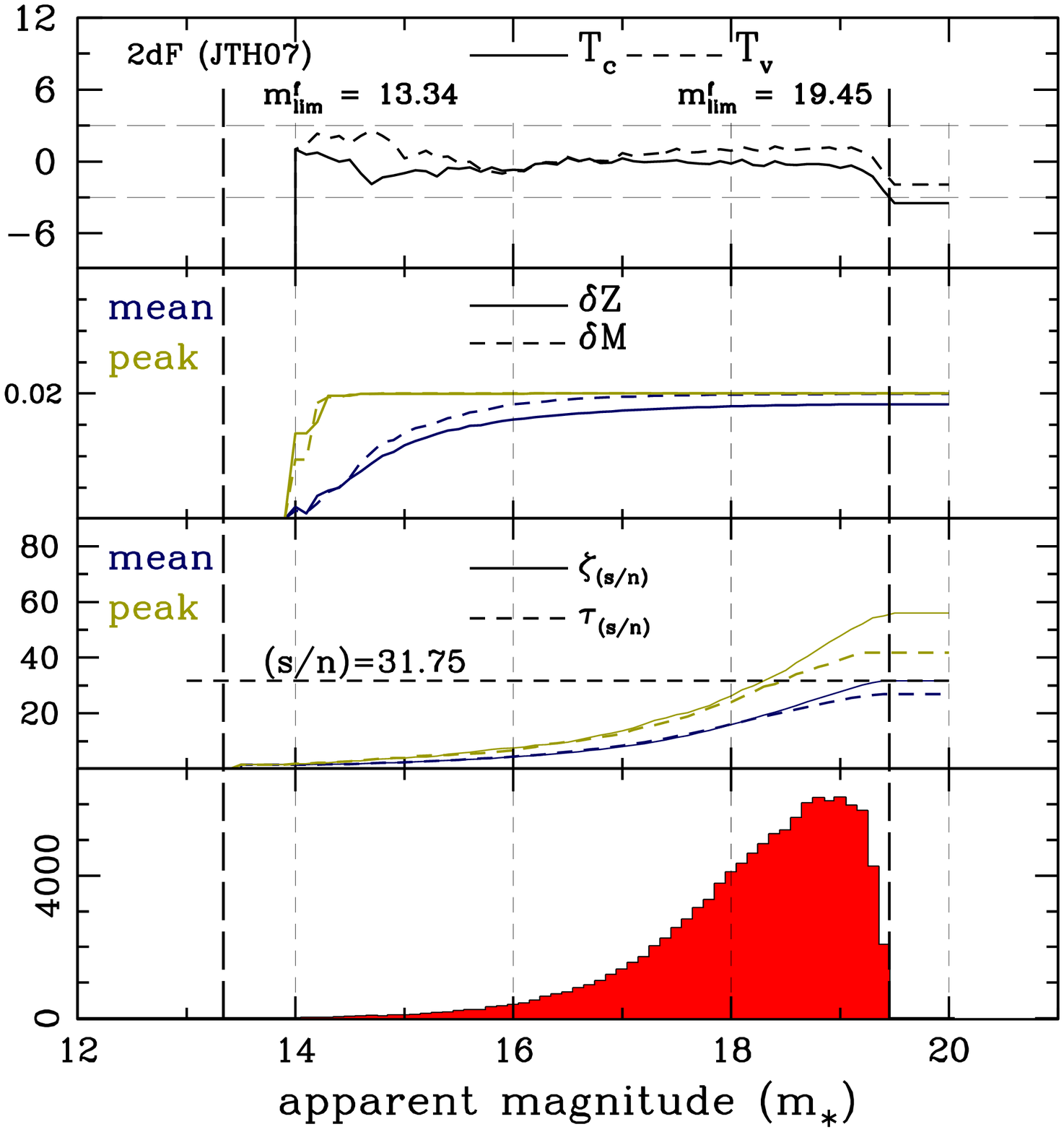}
      \caption{\small Plots demonstrating how we can establish a minimum signal-to-noise ({\sn}) threshold in SDSS-ET (left), MGC (middle) and 2dF (right) from the JTH07 approach.  For each survey we show the imposed apparent magnitude limits, {\mlimb} and {\mlimf}, as vertical black dashed lines.   In each case we choose a targeted, fixed value of {\deltazm} (2nd panels) for {\tctv} respectively (top panels) for which the resulting curves drop and `flat-line' on or below the $-3\sigma$ confidence limit of each test statistic. These widths correspond to {\deltazm}=0.015 (SDSS-ET), 0.065 (MGC) and 0.02 (2dF).  As these are our target  values, the 2nd panels for each survey show the resulting maximum (`peak'  shown in green) value of {\deltazm} that was achieved at each {\mstar}, as well as the mean values (shown in blue). To maximise the sampling range of apparent magnitude with the JTH07 method, we have included where necessary {\deltazm} widths {\it smaller} than the targeted, fixed value to be included in the calculation.  Therefore, it becomes clear that for the initial increments of {\mstar} where the distance between {\mstar} and {\mlimb} is small (coupled with low numbers of galaxies),  {\deltazm} will not reach the target width.   The mean values of {\deltazm} (shown in blue in the 2nd panels) for all surveys clearly illustrate this effect.  In the case of MGC, in particular, we observe initial increments of the peak curves indicating that for {\em no\/} galaxy are we able to construct a separable region with  {\deltazm} = 0.065.  By choosing the {\deltazm} widths to drop to the $-3\sigma$ limit beyond the {\mlimf}, we have, in effect, established a minimum {\sn} threshold that should ensure our estimators are not subject to the effects of very sparse sampling at brighter trial apparent magnitudes.  We find this choice corresponds to a {\sn}~$\sim$~12.3   for the SDSS data,  $\sim$~14.0 for MGC and  $\sim$~31.75 for the 2dF.  }
	\label{Fig:sn_thresh}
  \end{figure*}
\section{Implementation}\label{sec:implementation}%
\subsection{Establishing {\sn} Thresholds}\label{subsec:thresh}
With our expressions for the {\sn} of $\zeta_i$ and $\tau_i$ we now explore the way in which the concept of an {\sn} threshold, beyond which the {\tctv} statistics `flat-line', may be integrated into our code for computing these statistics for a given survey.  First we recall a fundamental property of both estimators, for a given {\mstar}:  by their construction, both {\tctv} should have a Gaussian sampling distribution with mean zero and a variance equal to unity.  We can therefore use the {\sn} expressions derived in the previous section to establish minimum {\sn} thresholds that will ensure the sampling distribution of both {\tctv} is indeed Gaussian, with the correct mean and variance, for each {\mstar} -- and particularly for fainter trial magnitudes closer to {\mlimf}.

A procedure by which we can achieve this is illustrated in {\F{Fig:sn_thresh}} and the discussion which follows.   In all three plots we present the following:
\begin{itemize}
\item{Top panel: the {\tctv} curves, shown as solid and dashed lines respectively, for a fixed, target value of {\deltazm} respectively.  We also indicate the imposed bright and faint apparent magnitude limits, {\mlimb} and {\mlimf} respectively.}
\item{2nd panel: here the achieved maximum (or peak) value of both {\deltazm}, at each {\mstar}, is shown in green, while the mean value of {\deltazm} is shown in blue.}
\item{3rd panel: here we show, for each {\mstar}, the resulting peak {\sn} indicated by the green curve, while the mean {\sn} is indicated by the blue curve.  In this case the solid lines represent the {\sn} for $\zeta_i$ whilst the dashed lines are for $\tau_i$.}
\item{4th panel: here we show a histogram of the apparent magnitude distribution for the survey under consideration.}
\end{itemize}

Let us consider the SDSS-ET survey, shown in the  left-hand plot of  {\F{Fig:sn_thresh}}.  Here we have applied the usual JTH07 method with a target width of {\deltazm} = 0.015.  We use the phrase `target width' as this version of the method seeks to maximise the sampling range of apparent magnitude within the JTH07 approach.  Thus,  we have  allowed {\deltazm} widths {\it smaller} than the targeted, fixed value to be included in the calculation. Therefore,  it becomes clear that for the initial increments of {\mstar} where the distance between {\mstar} and {\mlimb} is small (coupled with low numbers of galaxies),  {\deltazm} will not reach the target width.

In the 2nd panel of SDSS-ET, we show the resulting maximum value, as well as the mean value, of {\deltazm} that was achieved for each {\mstar}.  In particular the mean value curves are clearly seen to fall below the target width, as described above, for the initial increments of {\mstar}.

The choice of this width was made so that the {\tctv} curves drop to on or below their $-3\sigma$ confidence level at {\mlimf=17.45}.  In the case of SDSS-ET this value of {\deltazm} corresponds to a {\sn} level $\sim 12.3$ for both {\tctv}.  For this survey, therefore, one would need to maintain a minimum {\sn} threshold $\sim$12.3 to ensure that the {\tctv} statistics do not `flat-line' due to very sparse sampling at magnitudes brighter than {\mlimf=17.45}.  We will explore further the consequences of this in the following section.

In the remaining plots in {\F{Fig:sn_thresh}} we apply the corresponding procedure, with the same goal of ensuring that the `flat-lining' behaviour occurs for sufficiently small values of the test statistics, to the MGC and 2dFGRS.  For MGC in the middle plot, we require to set {\deltazm} = 0. 065 and find a mean {\sn}$\sim$14.0 threshold.  Finally, for 2dFGRS, we require to set {\deltazm} = 0.02, which corresponds to a mean threshold of {\sn} $\sim$ 31.75.
\subsection{Imposing the {\sn} Thresholds}
We can now use our pre-determined {\sn} levels, established in the previous section, and explore their impact on the {\tctv} estimators.  In {\F{Fig:sn_fnum}} we once again show the three surveys used for illustrative purposes in the same format as shown in {\F{Fig:sn_thresh}} and detailed in \S~\ref{subsec:thresh}.

Let us first consider the MGC data shown in the middle plot.  As a simplistic approach  to implementing an {\sn} threshold we have decided to keep the average {\sn} constant throughout the sampling procedure.   This is achieved by keeping constant the number of galaxies counted in $S_1\cup S_2$ (for $\zeta$) and $S_3\cup S_4$ (for $\tau$).   For MGC, to achieve the minimum {\sn} level of $\sim$14.0, already established  from {\F{Fig:sn_thresh}}, requires that the number of galaxies is equal to 150 in these combined regions.  If we look at the 2nd panel for MGC we can observe the consequences for both {\deltazm} as {\mstar} increases towards the true magnitude limit, {\mlimf} of the survey. Initially, we see that {\deltazm} are required to be rather large in size in order to achieve the minimum {\sn} level. This behaviour is expected and echoed by the histogram shown in the bottom panel of the plot.  As the density of  galaxies increases for fainter values of {\mstar} we see a sharp decline in the required width of {\deltazm} to achieve the same {\sn}.  We also note that imposing a minimum {\sn} level  restricts the magnitude sampling range within which  {\tctv} can reliably test completeness, particularly for brighter apparent magnitudes, and effectively introduces a value of {\mstar} at which the test statistics `initialise'.  In MGC, this initialisation occurs at an {\mstar}$\sim$17.6~mag.

If we now turn our attention to SDSS-ET on the left plot of  {\F{Fig:sn_fnum}} we can see that the distribution of galaxies on the {\mz} plane is such that we do not throw away much information on bright end of the apparent magnitude range.  Both {\tctv} initialise at around {\mstar}$\sim$15.1, after which we see a similar, steep drop in {\deltazm} as was apparent with MGC.  To achieve the minimum {\sn} level of $\sim 12.3$ the number of galaxies to be counted in separable regions is required to be 130 galaxies.

It is interesting to note that with the SDSS-ET survey, the {\tctv} statistics initially fluctuate below $-3\sigma$ between 15.1$\lesssim$ {\mstar} $\lesssim$ 15.5. Similar behaviour is also observed with 2dF (see the right-hand plot).  We recall that both SDSS-ET and 2dF  surveys are well described by a bright and faint apparent magnitude limit, and as such are subject to natural restrictions of the maximum size of the $\zeta$ and $\tau$ sample regions that retain the separability assumptions of the estimators - see \S~\ref{sec:currentissues} for further clarification of this point.  In our implementation of an {\sn} threshold to our code, we have in this instance, allowed {\deltazm} to grow in size beyond the limit imposed at the bright end.  Therefore, until {\deltazm} narrow to a width  that defines the separable region within the survey limits, the estimators will indicate incompleteness.  As we have already discussed, MGC can be well described by a faint limit only and is therefore not adversely affected by large values of {\deltazm}.

Finally, with the 2dF survey on right-hand panel of  {\F{Fig:sn_fnum}}, we have set the number of galaxies to 900 which seems to satisify our {\sn} criterion in our new scenario.  As we have just discussed, there are slight fluctuations below $-3\sigma$ at bright values of {\mstar}, i.e. for {\mstar}$\sim$16.4~mag. These correspond to the adoption of large widths for {\deltazm}.  It should be noted that even with the {\it minimum} {\sn} value, one can anticipate the true faint limit, {\mlimf}, of 2dF being being identified as brighter than the published limit of {\mlimf}=19.45 if one were to move to higher {\sn} levels.
\begin{figure*}\label{Fig:sn_fnum}
     \includegraphics[width=0.33\textwidth]{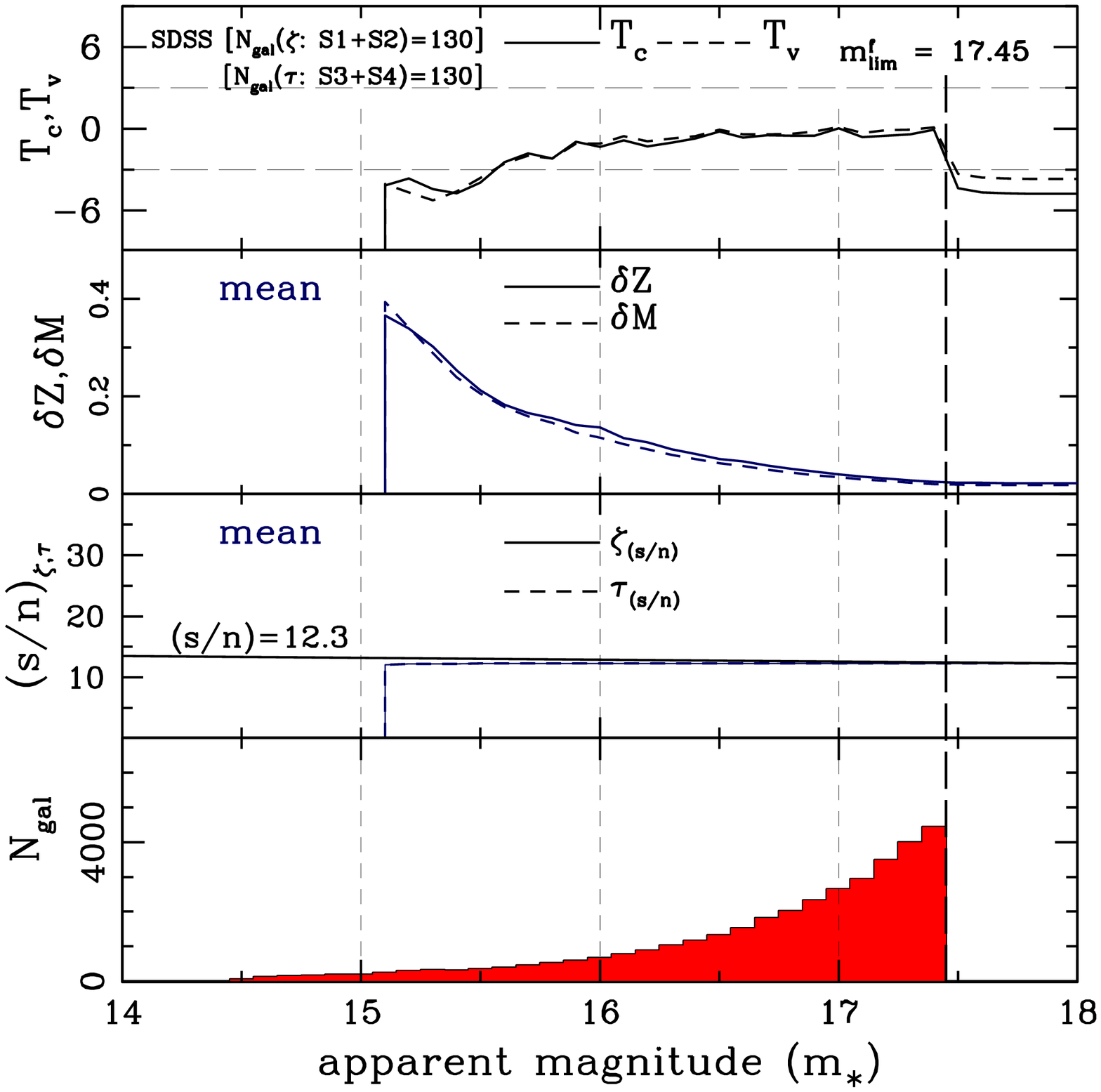}
     \includegraphics[width=0.33\textwidth]{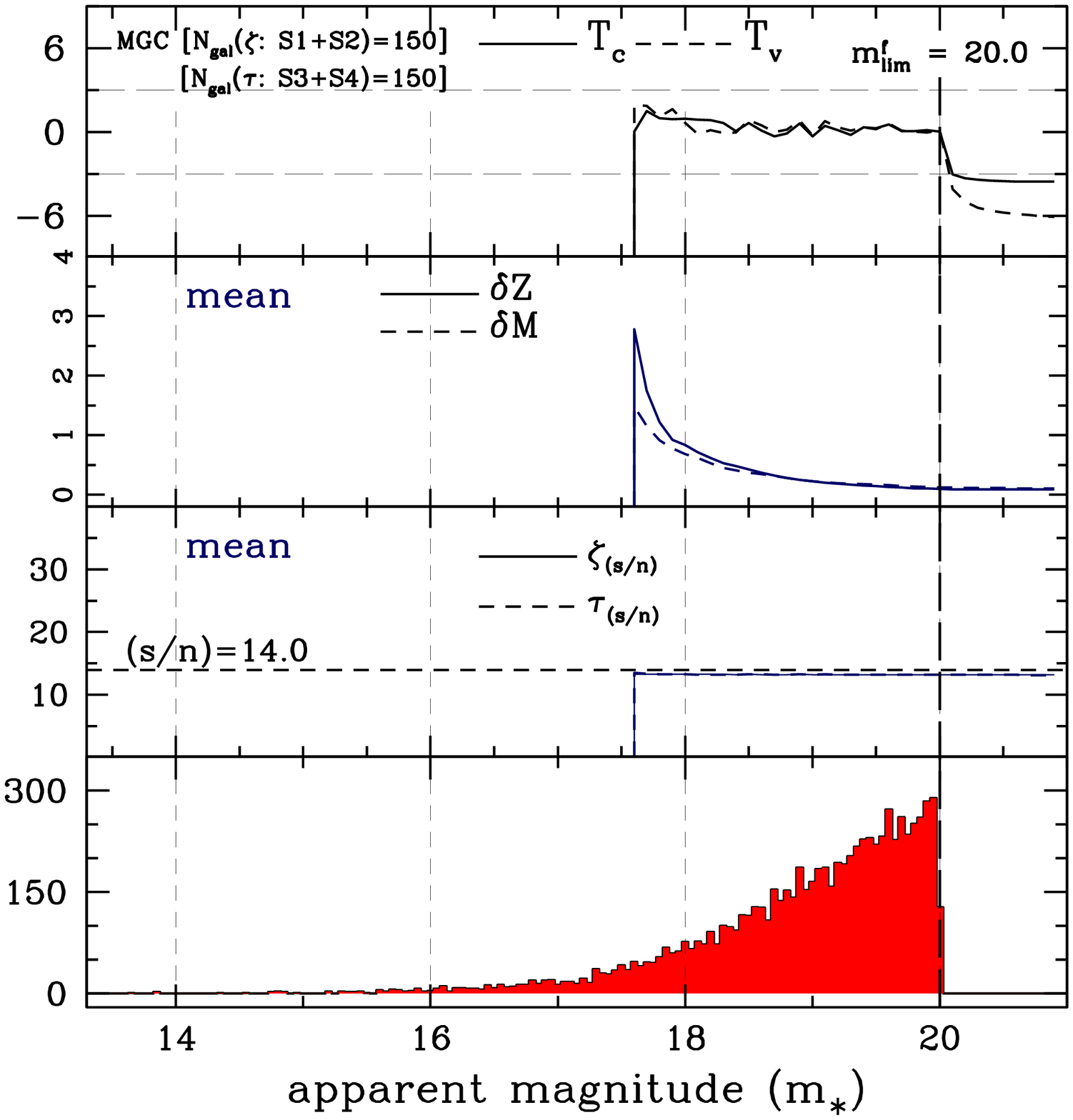}
    \includegraphics[width=0.33\textwidth]{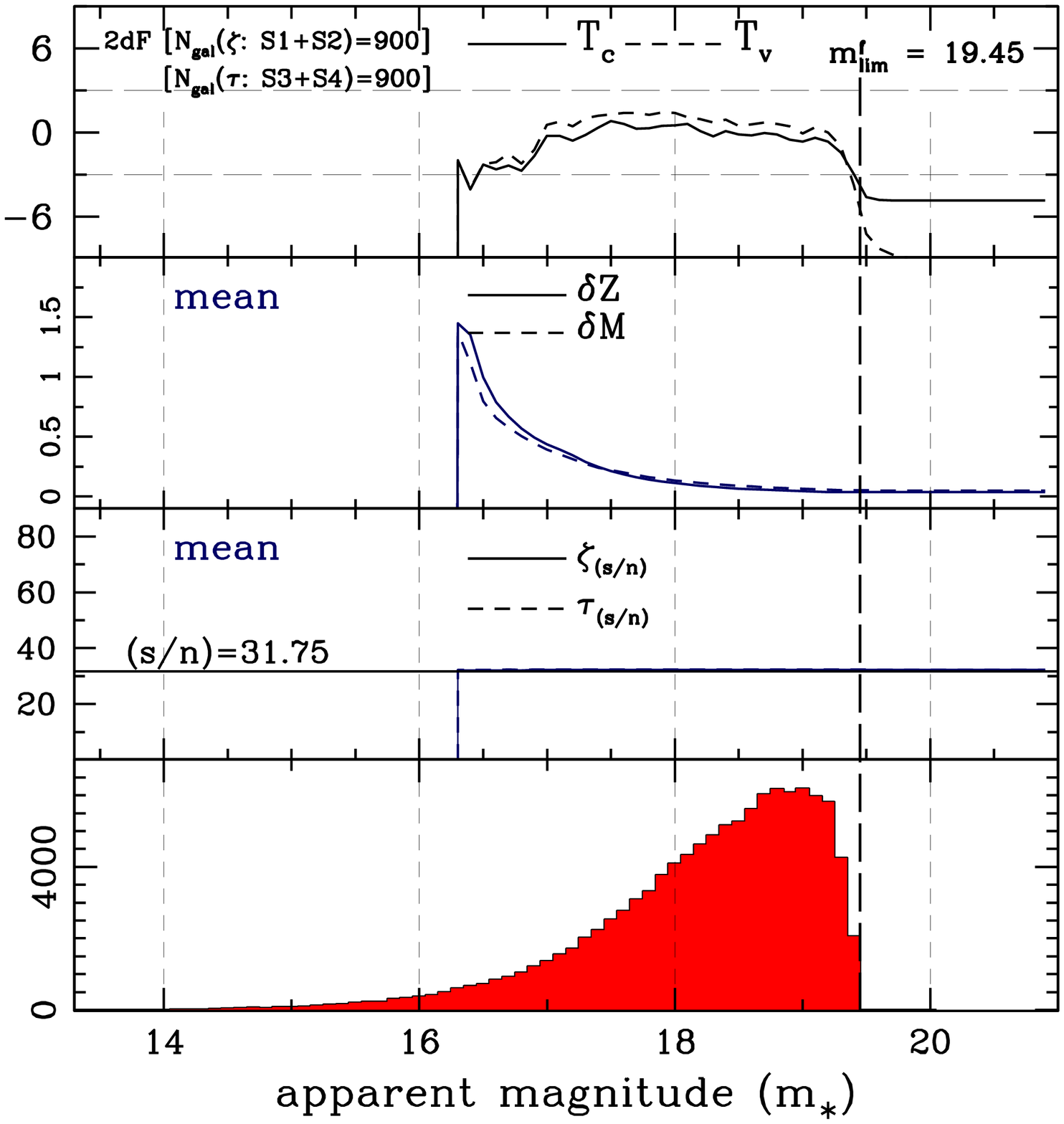}
      \caption{\small This figure shows the resulting {\tctv} curves for all three surveys when we adopt a constant {\sn} level based on the approximate thresholds as established in  {\F{Fig:sn_thresh}}. The first distinguishing feature from that presented in  {\F{Fig:sn_thresh}} is the deliberate omission of an imposed bright limit, {\mlimb}.  Secondly,  in this simplified approach we can keep the average {\sn} constant by keeping constant the number of galaxies we count in the $S_1\cup S_2$ (for $\zeta$) and $S_3\cup S_4$ for ($\tau$).   By doing this, however, of course we sacrifice information at the {\em bright\/} end of the luminosity function, where the survey is too sparsely sampled to be included in the calculation.  This effect is particularly obvious in the {\tctv} results for MGC (middle plot) and 2dFGRS (right plot)  and is mirrored in their respective histogram distributions.  However, the advantage of this approach is that we now have {\em adaptive\/} {\deltazm} widths for the respective estimators where, as we increase in {\mstar}, the {\mz} distribution is becoming more densely populated resulting in an almost asymptotic drop in the required widths to achieve the same {\sn} level.  As already mentioned, we have omitted {\mlimb}.   This allows us to achieve the minimum {\sn} for a greater range in apparent magnitude, and therefore  allows {\deltazm} to grow as large as required.   Such large values are only evident for initial values of {\mstar} as show in the 2nd panels of each survey.  For MGC (middle plot) this has no adverse effect on the respective {\tctv} curves as MGC is equally well described by a single faint apparent magnitude limit, {\mlimf}. However, we can see that for SDSS-ET  and 2dF, the {\tctv} statistics show initial fluctuations below $-3\sigma$ which is to be expected since both surveys are defined with bright magnitdue limits. As {\deltazm} decreases in size with increased {\mstar}, so the estimators become less sensitive to presence of {\mlimb}. }
	\label{Fig:sn_fnum}
  \end{figure*}
\section{discussion and concluding remarks}\label{sec:conclusions}
In this article we have introduced a method which attempts to optimize the completeness estimators, suitable for application to double-truncated galaxy survey data, as previously developed by \cite{Johnston:2007}. Our new approach resembles an ``adaptive smoothing" procedure which seeks to maintain a constant level of `information' -- as characterised by the signal-to-noise ratio computed for our test statistics -- allocated to each galaxy in the survey.  In applying this methodology to three well understood and characterized surveys, we have demonstrated the importance of properly accounting for the impact of sparse sampling in each galaxy survey. Furthermore, our results indicate that -- without adopting such a procedure -- the testing of magnitude completeness way be compromised, and spurious values for the `true' apparent magnitude limit(s) may be inferred. Thus, sparse sampling effects may impact adversely on previous applications of product-limit estimators which have been carried out in the literature to doubly-truncated data sets e.g. \cite{Efron:1999}.

The current article is the first of a two-part story. In the current paper we have set out to optimise our completeness estimators by imposing a lower limit on the number of galaxies contained in (and hence a lower limit on the width of) the rectangular regions we identify in the $M$-$Z$ distribution of our data. This lower limit ensures that the Gaussian sampling distribution, with mean zero and variance unity, of our {\tctv} statistics is preserved over the range of $m_*$ where the optimization is possible. In an upcoming publication (Johnston {\it et al} 2010, in preparation) we will consider in more detail the practical
implementation of these optimised estimators -- and in particular how we may use them to assign error bars to {\tctv}, and hence to compute confidence limits for the faint apparent magnitude limit, {\mlimf}, properly accounting for the correlations in {\tctv} between negihbouring values of the trial magnitude limit $m_*$.

%
%
%
\section*{Acknowledgements}
RJ would like to thank David Valls-Gabaud for his insightful comments and also the funding bodies EPSRC (UK) and the National Research Foundation (South Africa).

The Millennium Galaxy Catalogue consists of imaging data from the
Isaac Newton Telescope and spectroscopic data from the Anglo
Australian Telescope, the ANU 2.3m, the ESO New Technology Telescope,
the Telescopio Nazionale Galileo and the Gemini North Telescope. The
survey has been supported through grants from the Particle Physics and
Astronomy Research Council (UK) and the Australian Research Council
(AUS). The data and data products are publicly available from
http://www.eso.org/~jliske/mgc/ or on request from J. Liske or
S.P. Driver.

The SDSS-ET data-set was kindly provided by Mariangela Bernardi and Ravi Sheth.  Funding for the creation and distribution of the SDSS Archive has been provided by the Alfred P. Sloan Foundation, the Participating Institutions, the National Aeronautics and Space Administration, the National Science Foundation, the U.S. Department of Energy, the Japanese Monbukagakusho, and the Max Planck Society. The SDSS Web site is \url{http://www.sdss.org}. The SDSS is managed by the Astrophysical Research Consortium (ARC) for the Participating Institutions. The Participating Institutions are the University of Chicago, Fermilab, the Institute for Advanced Study, the Japan Participation Group, the Johns Hopkins University, Los Alamos National Laboratory, the Max Planck Institute for Astronomy (MPIA), the Max Planck Institute for Astrophysics (MPA), New Mexico State University, the University of Pittsburgh, Princeton University, the United States Naval Observatory, and the University of Washington.


%
%
%
\setcounter{equation}{0}
\renewcommand{\theequation}{A-\arabic{equation}}
\setcounter{section}{0}
\renewcommand{\thesection}{A-\arabic{section}}
\setcounter{figure}{0}
\renewcommand{\thefigure}{A-\arabic{figure}}
\bibliography{bibliography}
\bibliographystyle{astron}
\label{lastpage}
\onecolumn
\end{document}